**Phenomenological theory of cuprate superconductivity**

Mario Rabinowitz[a] and T. McMullen[b]
[a]Electric Power Research Institute, and Armor Research
715 Lakemead Way, Redwood City, CA 94062-3922
E-mail: Mario715@earthlink.net

[b]Physics Dept., Virginia Commonwealth University, Richmond, VA 23284 USA

**Abstract**

Reasonably good agreement with the superconducting transition temperatures of the cuprate high-$T_c$ superconductors can be obtained on the basis of an approximate phenomenological theory. In this theory, two criteria are used to calculate the superconducting transition temperature. One is that the quantum wavelength is of the order of the electron-pair spacing. The other is that a very small fraction of the normal carriers exist as Cooper pairs at $T_c$. The resulting simple equation for $T_c$ contains only two parameters: the normal carrier density and effective mass. We calculate specific transition temperatures for twelve cuprate superconductors.

**Introduction**

As the temperature of an ideal Bose gas is lowered, the particles undergo a Bose-Einstein (B-E) condensation when the thermal wavelength

$$\lambda_T = h / \left[2\pi m_B k_\beta T_c\right]^{1/2} , \qquad (1)$$

is comparable to the interparticle spacing [1,2]. Here h is Planck's constant, $k_\beta$ is the Boltzmann constant, $m_B$ is the Boson mass, and $T_c$ is the transition temperature. $\lambda_T$ is called the thermal wavelength, because $\lambda_T$ ♠ $\lambda$, the de Broglie wavelength.

For a three-dimensional ideal Bose gas, the B-E condensation temperature [1,2] is:

$$T_c^{BE} = \frac{h^2 n^{2/3}}{2\pi m_B k_\beta \left[\zeta(\tfrac{3}{2})\right]^{2/3}} = \frac{h^2 n^{2/3}}{11.92 m_B k_\beta} , \qquad (2)$$

where n is the three dimensional number density, $\zeta(x)$ is the Riemann zeta function of x, and $\zeta(3/2) = 2.612...$ . There are two shortcomings related to the ordinary B-E condensation. One is that $T_c^{BE}$ is orders of magnitude higher than the experimental $T_c$ of three-dimensional superconductors like the metallics if a sizable fraction of the normal carriers are bosons. The other is that the B-E gas does not condense in two dimensions so that unmodified B-E statistics are not a sufficient condition for B-E condensation. More general considerations argue against a one-dimensional condensation. However, with an energy gap (binding energy between particles), a B-E type condensation in lower dimensions is described by Blatt [3], although this appears to be contrary to Hohenberg's theorem[4].

Two conditions are met by known superconductors. One is the existence of bosons. The other is a condensation in momentum space. This led to a consideration that there may be two temperatures: $T_p$, the pairing temperature to form bosons out of fermion pairs; and the condensation temperature $T_C$. If $T_p \leq T_C$, then $T_p$ is the limiting temperature for superconductivity which we call the transition temperature $T_c$ . However it is difficult to calculate $T_p$ because this requires knowledge of the pairing interaction and its strength, although this is the usual strategy. Three possibilities were considered first by Rabinowitz [5-11] in 1987:

1. That $T_C$ might be less than or equal to $T_p$, in which case $T_c = T_C$.

2. That a small fraction of fermion pairs, $\sim k_\beta T_c/E_F$ exist above $T_c$.

(Since this fraction is so small, it would hardly change measurements of penetration depth and coherence length.)

3. Reduced dimensionality effects properly enter into λ via the equipartition principle.

His paradigm led to good agreement with experiment for broad classes of superconducting materials from 1 K to $10^9$ K. [5-9] More recently this has been extended to include $10^{-3}$ K for the superfluid state of $^3$He for a range of 12 orders of magnitude.

Even if $T_C \geq T_p$, the $T_C$ is found to be a much better upper limit to $T_c$ than $T_c^{BE}$. We are not aware of any $T_c$ calculations for specific cuprates, nor of any other attempts to calculate a reduced condensation temperature which does not include interactions. Here we present a more detailed test of this approach by comparing its results with the observed $T_c$'s of specific cuprates. Harshman and Mills[12] have compiled a comprehensive table of the relevant parameters, extracted from experimental data, which makes this possible. B-E condensation approaches to superconductivity were first presented by Schafroth, Butler, and Blatt [13,14], and more recently by Friedberg and Lee [15]. These approaches differ from ours, and appear not to have been successful in predicting $T_c$'s. Standard approaches have been unsuccessful with the cuprates.

The point of view adopted here is that as the temperature is decreased in a gas similar to a Bose-Einstein (B-E) gas, particles should start a B-E type condensation into the superconducting state when the de Broglie wavelength is of the order of the interparticle spacing for the small fraction of carriers that have become Cooper pairs [5-11].Thus we can avoid the difficult calculations encountered when the nature of the interaction is known, and which are impossible to do when the pairing interaction is not known. Here we use the previous approach of Rabinowitz for specific cases to calculate the transition temperatures for a dozen cuprate superconductors.

**Analysis**

We assume that a B-E type condensation in three dimensions of the free Bose gas involves electron pairs within a shell of energy ~ $k_\beta T_c$ around the Fermi energy, $E_F$. The number of electrons involved is the number of k values within the shell $\Delta k$ of energy width ~ $k_\beta T_c$ within the Fermi surface, so that $\Delta k$ is given by:

$$k_\beta T_c \approx \frac{\hbar^2 k_F^2}{2m} - \frac{\hbar^2 (k_F - \Delta k)^2}{2m}, \qquad (3)$$

where k is the wave vector, and m is the effective mass of the electron. Thus

$$\frac{\Delta k}{k_F} = \frac{k_\beta T_c}{2\left(\frac{\hbar^2}{2m} k_F^2\right)} = \frac{k_\beta T_c}{2E_F} . \qquad (4)$$

The number of bosons (Cooper pairs) for a roughly spherical Fermi surface is then

$$n_B = \tfrac{1}{2} \frac{\Delta k \left(4\pi k_F^2\right)}{\tfrac{4}{3}\pi k_F^3} = \frac{3\Delta k}{2k_F} n . \qquad (5)$$

Combining eqs. (4) and (5), we have

$$n_B = \frac{3kT_{c3}}{4E_F} n , \qquad (6)$$

where we introduce $T_{c3}$ , a "three-dimensional" transition temperature that may apply to some less anisotropic materials, and which can be thought of here as providing the energy cutoff that selects the effective bosons. The procedure will be to use $T_{c3}$ to determine the inter-boson spacing in order to calculate $T_{c2}$, which will be our transition temperature estimate for the cuprates. The choice of "two-dimensional" motion for these materials is based on their layered structure and highly anisotropic normal state properties.

Two electron pairs are encompassed by (1/2)λ when

$$\lambda \approx 2(n_B)^{-1/3} . \tag{7}$$

For a pair of electrons of effective mass 2m, momentum p, and kinetic energy $(f/2)k_\beta T$, the de Broglie wavelength is

$$\lambda = \frac{h}{p} = \frac{h}{\left[2(2m)(\frac{f}{2}k_\beta T_c)\right]^{1/2}} , \tag{8}$$

where f is the number of degrees of freedom per particle pair. For three-dimensional motion of the bosons we take f = 3. For two-dimensional motion we will take f = 2. Rabinowitz [7] gives a more general discussion of f.

Combining eqs. (6), (7), and (8) we obtain

$$T_{c3} = 0.218\left[\frac{\hbar^2 n^{2/3}}{2mk_\beta}\right] . \tag{9}$$

To obtain the "two-dimensional" transition temperature, λ as given by eq. (7) is equated to λ as given by eq. (8) with f = 2, as we restrict the kinetic energy to two dimensions by the equipartition of energy principle. This yields the two-dimensional transition temperature which we will compare with experiment:

$$T_{c2} = 0.328\left[\frac{\hbar^2 n^{2/3}}{2mk_\beta}\right] . \tag{10}$$

Table 1 shows a comparison of the experimentally observed transition temperatures, $T_c^{exp}$ with $T_{c2}$, together with the relevant input data as obtained from Harshman and Mills [12]. In eq. (10) we use the three-dimensional carrier density $n_{3D}$ for n, and the effective mass for m, where $m_o$ is the free electron mass.

**Conclusion**

It is remarkable how well our calculated transition temperatures agree with the experimental values without explicitly introducing a pairing mechanism. Perhaps this should not be entirely surprising, as $T_c$ itself is a measure of the interaction strength. In this model, $T_c$ enters into the equations in two different ways so that it is possible to solve for $T_c$. This may be why we can obtain the transition temperature without prior knowledge of the interaction mechanism.

**References**

1. D. S. Betts, Contemporary Physics 10 (1969) 241 .
2. K. Huang, Statistical Mechanics (John Wiley & Sons, Inc. N. Y., 1963).
3. J. M. Blatt, Theory of Superconductivity (Academic Press, N.Y. 1964).
4. P. C. Hohenberg, Phys. Rev. 158 (1967) 383 .
5. M. Rabinowitz, In Proceedings: EPRI Workshop on High-Temperature Superconductivity. pp. 2-3 to 2-33 (1987). EL/ER-5894
6. M. Rabinowitz, In Proceedings: EPRI Conference on Electrical Applications of Superconductivity. pp. 1-1 to 1-18 (1988). EL-6325-D
7. M. Rabinowitz, Intl. Journal of Theoretical Physics 28 (1989) 137 .
8. M. Rabinowitz, Physica C162-164 (1989) 249 .
9. M. Rabinowitz, Advances in Cryogenic Engineering (Plenum Press, New York. 1990) **36A**, 21.
10. M. Rabinowitz, Intl. Journal of Theoretical Physics , Basic Connection between Superconductivity and Superfluidity, to be published.
11. M. Rabinowitz, Appl. Phys. Commun., $^3$He is a Clear Indicator of a Common Basis for Superconductivity and Superfluidity, to be published.
12. D. R. Harshman and A. P. Mills, Phys. Rev. B 45 (1992) 10684 .
13. M.R. Schafroth, Phys. Rev. 100 (1955) 463 .
14. M.R. Schafroth, S.T. Butler, and J.M. Blatt Helv. Phys. Acta. 30 (1957) 93 .

15. R. Friedberg and T. D. Lee, Phys. Rev. B 40 (1989) 6745 .

TABLE 1. Comparison of Experimental and Theoretical Cuprate Transition Temperatures

| No. | Compound | $T_{exp}$ , K | $T_{c2}$, K | n, $10^{21}/cm^3$ | $m/m_o$ |
|---|---|---|---|---|---|
| 1 | $La_{1.9}Sr_{0.1}CuO_4$ | 33 | 46 | ~0.5 | ~2 |
| 2 | $La_{1.875}Sr_{0.125}CuO_4$ | 36 | 55 | ~1.0 | ~2.6 |
| 3 | $La_{1.85}Sr_{0.15}CuO_4$ | 39 | 51 | 5.2 | 8.6 |
| 4 | $YBa_2Cu_3O_{6.67}$ | 60 | 60 | 1.1 | 2.6 |
| 5 | $YBa_2Cu_3O_7$ | 92 | 80 | 16.9 | 12 |
| 6 | $YBa_2Cu_4O_8$ | 80 | 76 | 2.8 | 3.8 |
| 7 | $HoBa_2Cu_4O_8$ | 80 | 145 | ~1.3 | ~1.2 |
| 8 | $Bi_2Sr_2CaCu_2O_8$ | 89 | 43 | 3.5 | 7.8 |
| 9 | $(Bi_{1.6}Pb_{0.4})Sr_2Ca_2Cu_3O_{10}$ | 107 | 42 | 3.5 | 7.8 |
| 10 | $Tl_2Ba_2CaCu_2O_8$ | 99 | 50 | 4.9 | 8.4 |
| 11 | $Tl_2Ca_2Ba_2Cu_3O_{10}$ | 125 | 67 | 4.2 | 5.7 |
| 12 | $(Tl_{0.5}Pb_{0.5})Sr_2CaCu_2O_7$ | 80 | 90 | 2.8 | 3.2 |